\documentclass[aps,pra,showpacs,showkeywords,onecolumn,superscriptaddress]{revtex4-1}

\usepackage{graphicx, graphics}% Include figure files
\usepackage{mathcomp}
\usepackage{psfrag}
\usepackage{amsmath,amssymb,lmodern}
\usepackage{dcolumn}% Align table columns on decimal point
\usepackage{bm}% bold math
\usepackage[below]{placeins}
\usepackage{array}
\usepackage[usenames]{color}
\usepackage{textcomp}
\begin{document}

\title{Phase diagrams and multistep condensations of spin-1 bosonic gases in optical lattices}

\author{Xiaolei Zan}
\affiliation{Department of Physics, National University of Defense Technology, Changsha 410073, P. R. China}
\author{Jing Liu}
\affiliation{Department of Physics, National University of Defense Technology, Changsha 410073, P. R. China}
\author{Jianhua Wu}
\email{wujh@nudt.edu.cn}
\affiliation{Department of Physics, National University of Defense Technology, Changsha 410073, P. R. China}
\author{Yongqiang Li}
\email{li\_yq@nudt.edu.cn}
\affiliation{Department of Physics, National University of Defense Technology, Changsha 410073, P. R. China}

%\keywords{Keyword1, Keyword2, Keyword3}
%\linenumbers
\begin{abstract}
Motivated by recent experimental processes, we systemically investigate strongly correlated spin-1 ultracold bosons trapped in a three-dimensional optical lattice in the presence of an external magnetic field. Based on a recently developed bosonic dynamical mean-field theory (BDMFT), we map out complete phase diagrams of the system for both antiferromagnetic and ferromagnetic interactions, where various phases are found as a result of the interplay of spin-dependent interaction and quadratic Zeeman energy. For antiferromagnetic interactions, the system demonstrates competing magnetic orders, including nematic, spin-singlet and ferromagnetic insulating phase, depending on longitudinal magnetization, whereas, for ferromagnetic case, a ferromagnetic-to-nematic-insulating phase transition is observed for small quadratic Zeeman energy, and the insulating phase demonstrates the nematic order for large Zeeman energy. Interestingly, at low magnetic field and finite temperature, we find an abnormal multi-step condensation of the strongly correlated superfluid, i.e. the critical condensing temperature of the $m_F=-1$ component with antiferromagnetic interactions demonstrates an increase with longitudinal magnetization, while, for ferromagnetic case, the Zeeman component $m_F = 0$ demonstrates a local minimum for the critical condensing temperature, in contrast to weakly interacting cases.
\end{abstract}

%\flushbottom
\maketitle
%\thispagestyle{empty}

%\noindent Please note: Abbreviations should be introduced at the first mention in the main text – no abbreviations lists. Suggested structure of main text (not enforced) is provided below.

\section*{Introduction}
Spin-1 bosons, the simplest nontrivial spinor bosonic system, play an important role in understanding quantum magnetism~\cite{WU2006,Lewenstein2007,KAWAGUCHI2012253,RevModPhys.85.1191,KRUTITSKY20161,CAPPONI201650}. Due to the development of optical cooling and trapping techniques, ultracold quantum gases have provided an excellent laboratory for investigating many-body quantum systems with an unprecedented level of precision, and ultracold spinor gases have been realized for both bosons \cite{Stenger1998Spin,Barrett2001,Schmaljohann2004,Higbie2005,Zhao2015} and fermions \cite{Taie2010,DeSalvo2010,Lompe940,Zhang1467,PEbling2014,Pagano2014A,Scazza2014Observation,Hofrichter2016}. Experimental progresses motivate
theoretical studies on spinor bosonic gases, which manifest a range of phenomena absent in scalar Bose-Einstein condensates (BECs)~\cite{Batrouni2009,Soltanpanahi2010Multi,Heinze2011,Alexander2013,Mahmud2013,Natu2015,Zhu2017}. %Combinations of experimental and theoretical studies shed insight on a range of topics such as spin mixing~\cite{Law1998,Pu1999}, spin dynamics~\cite{Chang2004,Schmaljohann2004,Chang2005Coherent,Widera2005}, and phase transitions~\cite{Jacob2012,Zhao2015}.
Normally, the spinor system is prepared in a weak magnetic field to avoid blocking spin-exchange processes in the experimental timescale, and in parallel theoretical studies mainly focus on rich phase diagrams of spinor gases at zero magnetic field, where numerical exact methods, including quantum Monte-Carlo (QMC) or density-matrix renormalization group simulations, are mainly developed for low-dimensional cases~\cite{PhysRevLett.95.240404,PhysRevA.74.053419,PhysRevA.74.035601,Batrouni2009,PhysRevB.88.104509}. Actually, magnetic
field lifts the degeneracy of the ground states, and the interplay of spin-dependent and Zeeman interactions enriches strongly-correlated quantum phases~\cite{WU2006,Lewenstein2007,KAWAGUCHI2012253,RevModPhys.85.1191,KRUTITSKY20161,CAPPONI201650,Mohamed2013,Mobarak2013Tuning,PhysRevA.90.043609,PhysRevA.97.023628}.

Magnetism in ultracold atomic systems may be ascribed generally to local onsite interactions, and one can load the spinor Bose gas into an optical lattice so as to increase the low-energy density of states and enhance the role of interactions~\cite{WU2006,Lewenstein2007,KAWAGUCHI2012253,RevModPhys.85.1191,KRUTITSKY20161,CAPPONI201650}. Actually, one indeed tunes the spin-dependent interactions to the strength of the order of external Zeeman interactions (or microwave dressing field interactions)~\cite{Jiang2016First}, and study strongly correlated phenomena of the multiple spin states. Recently, a antiferromagnetic spinor condensate has been experimentally realized in a two-dimensional optical lattice, where, in a sufficiently deep lattice, a phase transition from a longitudinal polar phase to a broken-axisymmetry phase has been observed in steady states of spinor condensates~\cite{Zhao2015}. In a following-up experiment, they demonstrate evidence of first-order superfluid-Mott-insulating phase transitions in a lattice-confined antiferromagnetic spinor Bose-Einstein condensate~\cite{Jiang2016First}. These experiments on lattice systems have deepened the understanding of spinor systems and observed new phenomena which have not been predicted in theory~\cite{Uesugi2003,Svidzinsky2003,Mohamed2013}. Thus the influence of both the lattice setups and external magnetic fields should be reconsidered.

In contrast, the ferromagnetic spinor condensates have been studied less extensively~\cite{Imambekov,Tsuchiya,Kimura,Yamamoto}. Experimental studies on the ferromagnetic spin-1 bosons have been carried out mainly with $^{87}$Rb~\cite{Luo620}, which has a small spin-dependent interaction compared with the spin-independent interaction. Theoretical studies demonstrate that the system exhibits saturated ferromagnetism over the entire zero-temperature phase diagram in the absence of external magnetic fields~\cite{Katsura2013}, where properties of the system are similar to those of scalar bosons and the phase transitions are of second-order. However, the system presents rich phases in the presence of an external magnetic field, due to the competition between the quadratic Zeeman effect and the ferromagnetic interactions \cite{KAWAGUCHI2012253,RevModPhys.85.1191}. Recently, a theoretical research is carried out for ferromagnetic spin-1 gases under an external magnetic field, which indicates discontinuous first-order phase transitions~\cite{PhysRevA.96.023628}. Despite the rich and interesting results obtained above, there is still a lack of relevant researches about competing spin-ordered Mott phases.

Exploring the thermodynamics of interacting many-body systems has been arguably one of the most important achievements of cold-atomic gases. It is an interesting topic examining the connection between magnetic orders and Bose-Einstein condensates of ultracold multispecies atomic gases~\cite{WU2006,Lewenstein2007,KAWAGUCHI2012253,RevModPhys.85.1191,KRUTITSKY20161,CAPPONI201650}. For spinor gases, a multi-step condensation has been predicted theoretically~\cite{Isoshima2000Double,zhang2003,zhang2004bose,Kis2006Phases,Phuc2011effect,Lang2014Therm} and observed experimentally~\cite{PhysRevLett.102.125301,PhysRevA.86.061601,PhysRevA.90.023610}. For small Zeeman field, for example, antiferromagnetic interactions qualitatively change the phase diagram and lead to condensation in $m_F\pm1$ state, which is a phenomenon that cannot occur for an ideal gas~\cite{Frapolli2017Stepwise}. As far as we know, however, there are still lack experiments on multistep condensations of the spinor bosonic gases in an optical lattice, and it is still unclear for condensing sequences of strongly correlated bosonic gases in optical lattices.

%Quite abundant researches has been carried out in the investigation of the spinor ultracold gases trapped in a periodic optical lattice with negative and positive on-site spin-dependent interactions in the presence of an external magnetic field, the dynamics of the system in a quantum quench scenario have been studied by mean-field approximation~\cite{Mahmud2013}, while the numerical exact approach, QMC, has been applied in a two-dimensional case~\cite{PhysRevB.88.104509} to obtain related phase diagrams. Quite recently, Standard Basis Operator (SBO) method~\cite{Zhu2017} has been occupied to investigate similar issues and the results obtained show good agreements with those calculated via QMC. In this paper, we employ bosonic dynamical mean-field theory to carry on further calculations.
In this paper, we focus on strongly correlated spinor ultracold gases in a three-dimensional (3D) optical lattice and systemically investigate the system for both negative and positive on-site spin-dependent interactions in the presence of an external Zeeman interaction, based on bosonic dynamical mean-field theory.
Our study here is an extensive calculations performed in Ref.~\cite{PhysRevA.93.033622}. For antiferromagnetic interactions, we take $^{23}$Na as examples, and determine the many-body phase diagrams, including nematic phase, ferromagnetic phase, spin-singlet insulator and different types of superfluid phase.
As for the ferromagnetic case, we take $^7$Li and $^{87}$Rb as examples, and map out the phase diagrams, including ferromagnetic and nematic insulating phase, and superfluid phase. Moreover, we study the stability of these quantum phases against thermal fluctuations, obtaining finite temperature phase diagrams. Interestingly, a multi-step condensation for the Zeeman components is explored, and an abnormal condensation observed, in contrast to weakly interaction cases.
\section*{Methods}
\subsection*{The Model}
We consider spin-1 bosonic gases with spin-dependent interactions loaded in an optical lattice. We assume that an external magnetic field is applied along the quantization axis. In sufficiently deep lattices and the single-mode approximation, the physics of the system is described by the extended Bose-Hubbard model ~\cite{RevModPhys.85.1191}:
\begin{equation}
\hat H=-t\sum_{\langle i,j\rangle,\sigma}(a^\dagger_{i,\sigma}a_{j,\sigma}+a^\dagger_{j,\sigma}a_{i,\sigma})+\frac{U_{0}}{2}\sum_{i}\hat n_{i}(\hat n_{i}-1)+\frac{U_2}{2}\sum_{i}(S^{2}_{i}-2\hat n_{i})+pS_{iz}+q\sum_{i,\sigma}\sigma^{2}\hat n_{i,\sigma} -\mu\sum_{i}\hat n_{i}
\label{eq:hubbard}
\end{equation}
where, $a_{i,\sigma}$ ($a^\dagger_{i,\sigma}$) denotes the annihilation (creation) operator of a boson at lattice site $i$, and spin $\sigma\in\{-1,0,1\}$. The particle-number operators are defined by $\hat n_{i,\sigma}:=a^\dagger_{i,\sigma}a_{i,\sigma}$ and $\hat n_{i}:=\sum_{\sigma}\hat n_{i,\sigma}$, the spin operators by $S^{(\alpha)}_{i}:=\sum_{\sigma,\sigma'}a^\dagger_{i,\sigma}S^{(\alpha)}_{\sigma,\sigma'}a_{i,\sigma'}$ for $\alpha=x,y,z$, where $S^{(\alpha)}_{\sigma,\sigma'}$ denotes the elements of spin-1 matrices $S^{(\alpha)}$.
$t$ is the hopping amplitude, and $\langle i,j\rangle$ means the sum over nearest neighbors.
The second term $U_{0}$ is the on-site interaction between atoms, and the third term $U_{2}$ represents spin-dependent interaction on the same site.
The coefficient $p$ denotes linear Zeeman energy (equivalent to magnetization $M_z$), $q$ quadratic Zeeman energy, and $\mu$ the chemical potential.
In the following we consider both antiferromagnetic $U_2/U_0>0$ and ferromagnetic interactions $U_2/U_0<0$. For instance, in experiments with $^{23}$Na atoms the spin-dependent interaction $U_2/U_0\simeq 0.037$~\cite{Zhao2015}, with $^7$Li atoms with $U_2/U_0 \simeq -0.7$~\cite{PhysRevA.68.063602}, and with $^{87}$Rb atoms with $U_2/U_0\simeq -0.005$~\cite{Luo620}.

\subsection*{Bosonic Dynamical Mean-Field Theory}
%\subsubsection{BDMFT equations}
To investigate quantum phases of spinor Bose gases loaded into a cubic optical lattice, described
by Eq.~(\ref{eq:hubbard}), we recently establish a bosonic version of dynamical mean-field theory on the generic three-dimensional situation, where details can be found in Ref.~\cite{PhysRevA.93.033622}. Here we only show the basic idea of this method.
As in fermionic dynamical mean-field theory, the main idea of our BDMFT approach is to map the quantum lattice problem with many degrees
of freedom onto a single site coupled self-consistently to a noninteracting bath~\cite{Georges1996Dynamical,Vollhart_08,Hubur_08,Hu2009Dynamical,Li2011Tunable,He2012Quantum,Li2013Lattice,Li2012Anisotropic}.
The dynamics at the impurity site can thus be thought of as the hybridization of this site with the bath. Therefore, properties of the many-body system can be captured by a single impurity model.
In a more formal language, Hamiltonian ~(\ref{eq:hubbard}) is mapped onto a single-site problem and described by a local effective action:
%\begin{small}
\begin{eqnarray}
S_{\mathrm{imp}}=&-&\int^{\beta}_{0}\!d\tau d\tau'\!\sum_{\sigma\sigma'}
\!\left(
\begin{array}{c}
 a^\ast_{0,\sigma}(\tau)\quad  a_{0,\sigma}(\tau)
\end{array}
\right)
{\bm {\mathcal{G}}}^{-1}_{0,\sigma\sigma'}
\!\left(
\begin{array}{c}
a_{0,\sigma'}(\tau')\\
a^\ast_{0,\sigma'}(\tau')\\
\end{array}
\right) +\int^{\beta}_{0}d\tau\{\frac{U_0}{2}n_{0}(\tau)(n_{0}(\tau)-1)+\frac{U_2}{2}({\bf{S}}^{2}_{0}(\tau)-2n_{0}(\tau)) \nonumber\\
&-&\,t\sum_{\langle0,i\rangle,\sigma}(a^\ast_{0,\sigma}(\tau)\phi_{i,\sigma}(\tau)+a_{0,\sigma}(\tau)\phi^\ast_{i,\sigma}(\tau))+\,p S_{iz}(\tau)+ q\sum_{i,\sigma}\sigma^{2} n_{i,\sigma}(\tau)\},
\label{eq:BDMFT}
\end{eqnarray}
%\end{small}
where $0$ is index of the impurity site, $\tau$ is the imaginary time, $\bm{\mathcal{G}}^{-1}_{0,\sigma\sigma'}(\tau-\tau')$
denotes a local non-interacting propagator interpreted as a local dynamical Weiss Green's function,
\begin{equation}
\bm{\mathcal{G}}^{-1}_{0,\sigma\sigma'}(\tau-\tau')\equiv-
\left(
\begin{array}{cc}
(\partial_{\tau'}-\mu_{\sigma})\delta_{\sigma\sigma'}+t^2\sum\limits_{\langle0i\rangle,\langle0j\rangle}G^{1}_{\sigma\sigma',ij}(\tau,\tau') & t^{2}\sum\limits_{\langle0i\rangle,\langle0j\rangle}G^{2}_{\sigma\sigma',ij}(\tau,\tau')\\
t^{2}\sum\limits_{\langle0i\rangle,\langle0j\rangle}G^{2\ast}_{\sigma\sigma',ij}(\tau',\tau) & (-\partial_{\tau'}-\mu_{\sigma})\delta_{\sigma\sigma'}+t^2\sum\limits_{\langle0i\rangle,\langle0j\rangle}G^{1}_{\sigma\sigma',ij}(\tau',\tau)\\
\end{array}
\right)
\label{eq:weiss}
\end{equation}
and the superfluid order parameters is given by $\phi_{i,\sigma}(\tau)\equiv\langle a_{i,\sigma(\tau)}\rangle_{0}$ with $\langle\cdots\rangle_{0}$ indicating that the expectation value is calculated in the cavity system\cite{Hubur_08,Li2011Tunable}.
We find that the effective action~(\ref{eq:BDMFT}) is represented by an Anderson impurity Hamiltonian
\begin{eqnarray}\label{eq:AndersonHamiltion}
\hat H_{A}=& &-t\sum_{\sigma}(\phi^{\ast}_{\sigma}\hat a_{0\sigma}+\mathrm{h.c.})+\frac{U_{0}}{2}\hat n_{0}(\hat n_{0}-1)+\frac{U_{2}}{2}(\bm{\hat{S}}^{2}_{0}-2\hat n_{0})+q\sum_{i,\sigma}\sigma^{2}\hat n_{i,\sigma}-\sum_{\sigma}\mu_{0\sigma}\hat n_{0\sigma}+\sum_{l}\epsilon_{l}\hat b^{\dagger}_{l}\hat b_{l} \\
& &+\sum_{l,\sigma}(V_{\sigma,l}\hat a_{0\sigma}\hat b^{\dagger}_{l}+W_{\sigma,l}\hat a_{0\sigma}\hat b_{l}+\mathrm{h.c.}). \nonumber
\end{eqnarray}
We remark three points here. First the chemical potential and interaction term are directly inherited from the Hubbard Hamiltonian. Second,
there are two baths coupling to the impurity site: the bath of condensed bosons is represented by the Gutzwiller term with superfluid order parameter $\phi_{\sigma}$ for each component, and the bath of normal bosons is described by a finite number of orbitals with creation operators $\hat b^{\dagger}_{l}$ and energies $\epsilon_{l}$, where these orbitals are coupled to the impurity via normal-hopping amplitudes $V_{\sigma,l}$ and anomalous-hopping amplitudes $W_{\sigma,l}$. Third, we employ the exact diagonalization (ED) solver for the Anderson impurity problem~\cite{Hubur_08}. The maximal number of normal bath orbitals in ED is limited to $n_s = 4$, and the Fock space for the impurity bosons is truncated at a maximum occupation number $n=6$. We systematically verified that the used maximal numbers in the present study are sufficient for ED.

By exact diagonalization of Ham.~(\ref{eq:AndersonHamiltion}), we obtain the local Green's function $\bm{G}_{\sigma\sigma'}(i\omega_{n})$, and the local self-energy is then obtained from the local Dyson equation
\begin{equation}
 \bm{\mathcal{G}}^{-1}_{\sigma\sigma'}(i\omega_{n})=(i\omega_{n}\sigma_{z}+\mu_{\sigma})\delta_{\sigma\sigma'}-\bm{\Delta}_{\sigma\sigma'}(i\omega_{n})=\bm{\Sigma}_{\sigma\sigma'}(i\omega_{n})+\bm{G}^{-1}_{\sigma\sigma'}(i\omega_{n}),
\label{eq:Weiss Green's function}
\end{equation}
where we have defined the hybridization functions:
\begin{eqnarray}
 \Delta^{1}_{\sigma\sigma'}(i\omega_{n})\equiv-\sum_{l}(\frac{V_{\sigma,l}V_{\sigma',l}}{\epsilon_{l}-i\omega_{n}}+\frac{W_{\sigma,l}W_{\sigma',l}}{\epsilon_{l}+i\omega_{n}}) &\nonumber\\
 \Delta^{2}_{\sigma\sigma'}(i\omega_{n})\equiv-\sum_{l}(\frac{V_{\sigma,l}W_{\sigma',l}}{\epsilon_{l}-i\omega_{n}}+\frac{W_{\sigma,l}V_{\sigma',l}}{\epsilon_{l}+i\omega_{n}}). &
\label{eq:hybridization}
\end{eqnarray} We approximate the lattice self-energy $\bm{\Sigma}_{\mathrm{lat},\sigma\sigma'}$ by the impurity self-energy $\bm{\Sigma}_{\sigma\sigma'}$, and the self-consistency loop is then completed by the conditions for lattice Green's function
\begin{equation}
\bm{G}_{\mathrm{lat}}(i\omega_{n})=\int \epsilon \rho(\epsilon)\frac{1}{i\omega_{n}\sigma_{z}+\mu_{\sigma}-\bm{\Sigma}(i\omega_{n})-\epsilon},
\label{eq:lattice Green's function}
\end{equation}
where $\rho(\epsilon)$ denotes the density of states for three-dimensional cubic lattices at energy $\epsilon$.
Note here that this method is exact for infinite dimensions, and is a reasonable approximation for high but finite dimensions, where the reliability of this method has been verified by comparison with quantum Monte-Carlo simulation~\cite{QMC_2007,Werner_12}. Our approach is a non-perturbative method, including the local quantum fluctuations of the strongly correlated systems exactly, but neglecting non-local spatial correlations.

For most of cases, the self-consistent BDMFT loop yields stable solutions, but, for some cases, it produces multiple stable results, especially around the phase boundary for first-order transition. To find the many-body ground states in these cases, we calculate the energy within BDMFT, and for the Bose-Hubbard model of spin-1 bosons, the local energy is given by:
\begin{eqnarray}
E=\Bigg[ \frac12\sum_i\Big\langle U_0 n_i(n_i-1) + U_2 ({\bf S}^2_i - 2n_i)+2\Big(pS_{iz} + q\sum_\sigma \sigma^2 n_\sigma\Big) \Big\rangle -k_BT\sum_{i,\sigma,n}\int d\epsilon\, \epsilon \rho(\epsilon) G_{i\sigma}(i\omega_n) - t\sum_{\langle ij \rangle \sigma} \phi^\ast_{i\sigma}\phi_{j\sigma} \Bigg] / N.
\end{eqnarray}
Here, $N$ is the total number of lattice sites, $i\omega_n=2n\pi/\beta$, and $G_\sigma(i\omega_n)$ denotes the local Green's function.

\section*{RESULTS}
\subsection*{Antiferromagnetic interactions} \label{subsec:ferromagnetic interactions}
In this section, we focus on experimentally accessible parameters for $^{23}{\rm Na}$, ${\it i.e.}$ $U_2/U_0=0.037$, in a 3D optical lattice constructed from a single-mode laser at wavelength $\lambda = 1064$ nm with recoil energy $E_R = 2h^2/m \lambda^2$. In this limit of spin-dependent interaction $U_2 \ll U_0$, a first superfluid-insulator transition featured by hysteresis effect and significant heating~\cite{Jiang2016First} is observed experimentally by changing the ratio $U_0/t$. Motivated by the recent experimental progresses, we investigate the spin-1 bosonic gases in a 3D optical lattice in the presence of an external magnetic field. Actually, the interplay of spin-dependent and Zeeman interactions in the system gives rise to an abundant phase diagram. At finite temperature, interestingly, the strongly correlated system may demonstrate an abnormal multi-step condensation for different species, due to quantum and thermal fluctuations.% We will discuss these issues in more detail in the following.
%\begin{figure}[t!]
%\centering
%\begin{tabular}{cc}
%\includegraphics*[width=0.515\linewidth]{n2.eps}
%\hspace{-4.5mm}
%\includegraphics*[width=0.515\linewidth]{n3.eps}
%\end{tabular}
%\vspace{-5mm}
%\caption{(Color online) Zero-temperature phase diagram in the Mott-insulating regime of the spin-1 ultracold bosons in a 3D cubic lattice ($V=20 E_R$) for antiferromagnetic interaction $U_2/U_0=0.037\;(\rm ^{23}Na)$, obtained via BDMFT. Our method resolves a filling-dependent magnetic spin order, including FM, NI and SSI. %In contrast to mean-field theories, these phases including SF and NI demonstrate non-zero local magnetization $M\neq0$, obtained by BDMFT. The Mott-insulating phase (MI) features filling-dependent spin structure: the MI with an odd number of atoms per site has spin $S\neq0$ (NI), whereas the MI with an even number of atoms shows competition between SSI and NI.
%}\label{Ne_1}
%\end{figure}
\begin{figure}[t!]
\centering
\begin{tabular}{c}
\includegraphics*[width=0.9\linewidth]{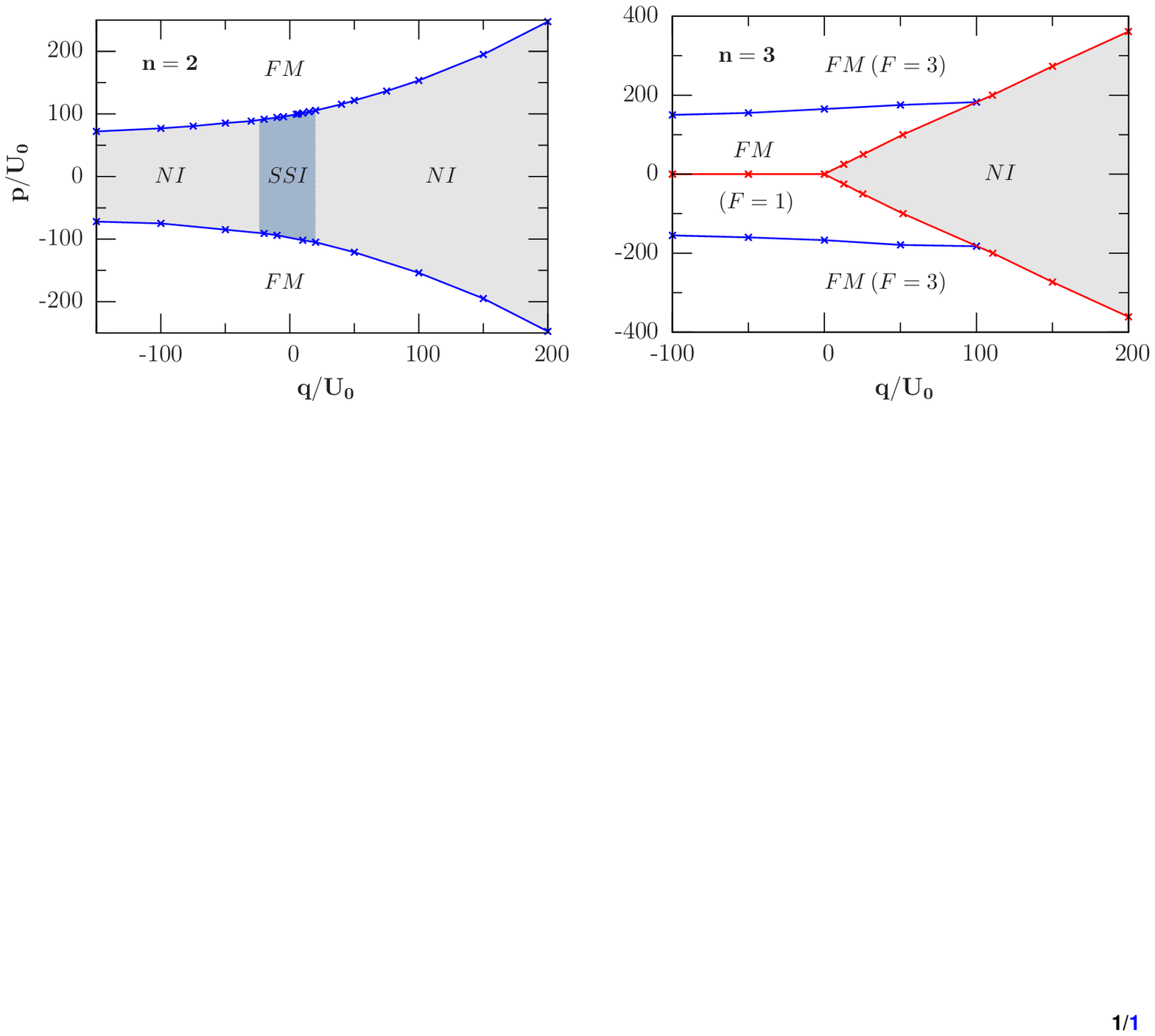}
\end{tabular}
\vspace{-3mm}
\caption{(Color online) Zero-temperature phase transition for spin-1 ultracold bosons in a 3D cubic lattice with an antiferromagnetic interaction $U_2/U_0=0.037$  for fixed filling $n=2$. }\label{Ne_1}
\end{figure}

\begin{figure}[t!]
\centering
\begin{tabular}{c}
\includegraphics*[width=0.6\linewidth]{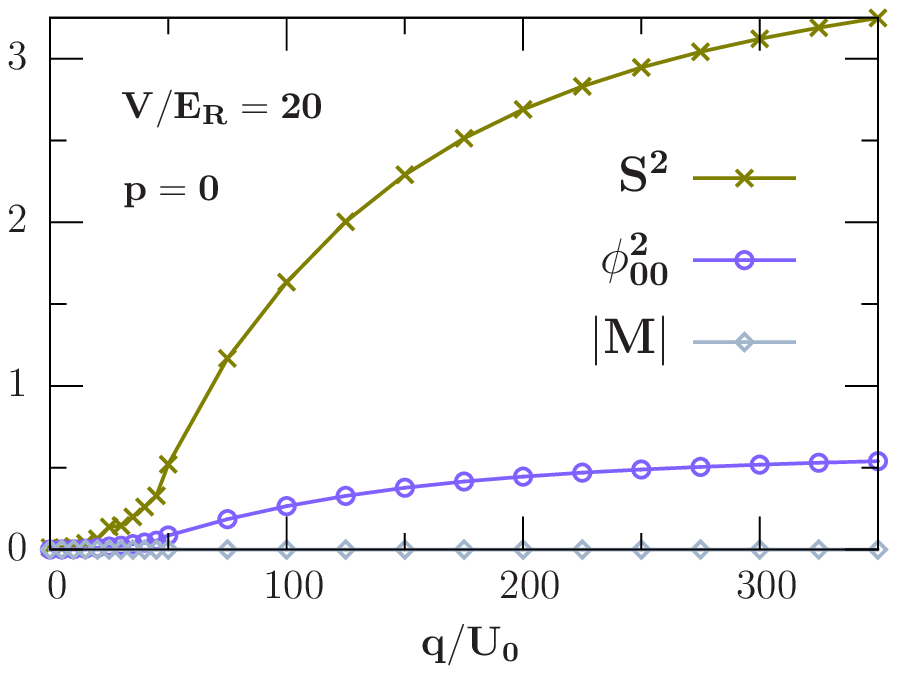}
\end{tabular}
\vspace{-3mm}
\caption{(Color online) Zero-temperature phase transition for spin-1 ultracold bosons in a 3D cubic lattice with an antiferromagnetic interaction $U_2/U_0=0.037$  for fixed filling $n=2$. }\label{transition}
\end{figure}
\subsubsection*{Zero temperature}
\paragraph{Phase diagram in deep Mott insulator.}

First, we study strongly correlated phases in the deep Mott insulator (MI) and resolve its long-range spin order in the presence of an external magnetic field. We summarized our calculations of the zero-temperature case in Fig.~\ref{Ne_1}, where two phase diagrams are shown in the $p$-$q$ plane with an antiferromagnetic interaction $U_2/U_0\approx0.037\;(\rm ^{23}Na)$ and filling $n=2$ (left) and $3$ (right), respectively. The depth of the optical lattice is chosen to be $V=20\, E_R$ (t/$U_{0} \approx 0.012$), where the system is in a typical Mott insulating regime, with $E_R$ being the recoil energy of atoms. Three distinct phases, namely ferromagnetic phase (FM), nematic insulator (NI) and spin-singlet insulator (SSI), are observed and we employ the values of the condensate order parameter $\phi^1_\alpha\equiv\langle b_\alpha\rangle$, the nematic order parameter $\phi^2_{\alpha\beta} \equiv \langle S_\alpha S_\beta \rangle -\delta_{\alpha\beta}/3 \langle S^2 \rangle$, and the local magnetization $\bf {M}\equiv\langle {\bf S} \rangle$ to characterize them respectively.

For filling $n=2$ and $V=20\, E_R$, our BDMFT calculations predict that different parameters will result in different Mott insulating phases with different types of spin order including FM featured by $\phi^1_\alpha =0 $, ${\bf M} \neq 0$, NI by $\phi^1_\alpha =0 $, $\phi^2_{\alpha\beta}>0$ and ${\bf M}=0$ and SSI by $\phi^1_\alpha =0$, $\phi^2_{\alpha\beta}=0$ and $\langle {\bf S}^2\rangle=0$ with $S$ being the local total spin (due to the symmetry of the spin wave-function on each site, $n+S=({\rm even})$~\cite{Imambekov}). It is worth mention that the formation of singlet pairs will give rise to an even number of atoms per site which characterize SSI phase. Actually, for the small linear Zeeman energy $p$ (equivalent to magnetization $M_z$) and quadratic Zeeman energy $q$, the system is in a regime dominated by spin-dependent interactions, and favors the SSI phase by forming a singlet pair and lowering the spin-dependent energy. For larger quadratic Zeeman energy $q$, the degeneracy of the three magnetic species is lifted, and the system goes through a phase transition from the SSI to the NI phase, $\it i.e.$ the system enters into the NI phase with $n_0 > n_{\pm1}$ for positive $q$, and with $n_{\pm1} > n_0$ for negative $q$, as shown in Fig.~\ref{transition} with $p=0$.  For larger $p$ (magnetization $M_z$), one species of $n_{\pm1}$ is energetically favored and the FM phase develops. We notice that the phase diagrams are symmetric upon linear Zeeman interactions and this symmetry is also manifested in the Hamiltonian~(\ref{eq:hubbard}).

We next move to three identical particles with $n=3$, and the corresponding phase diagram is shown in the right panel of Fig.~\ref{Ne_1}. For small magnetic field, the system favors the NI phase with $\langle S^2 \rangle =2$ by lowering spin angular momentum, where the detailed discussion can be found in Ref.~\cite{PhysRevA.93.033622}. For large quadratic Zeeman energy $q$, FM phase with $|{\bf M}| =3$ is energetically favored by populating the $n_1$ species for $p>0$ and by the $n_{-1}$ species for $p<0$, respectively. For intermediate $p$ and negative $q$, the system demonstrates a FM phase with $\langle S^2 \rangle =2$ and $|{\bf M}| =1$. Note here that, along the line $p\approx0$, the system favors the NI phase for both positive and negative quadratic Zeeman energies (see the red lines in the right panel of Fig.~\ref{Ne_1}).
\begin{figure*}[t!]
\centering
\vspace{-3mm}
\includegraphics[width=0.9\linewidth]{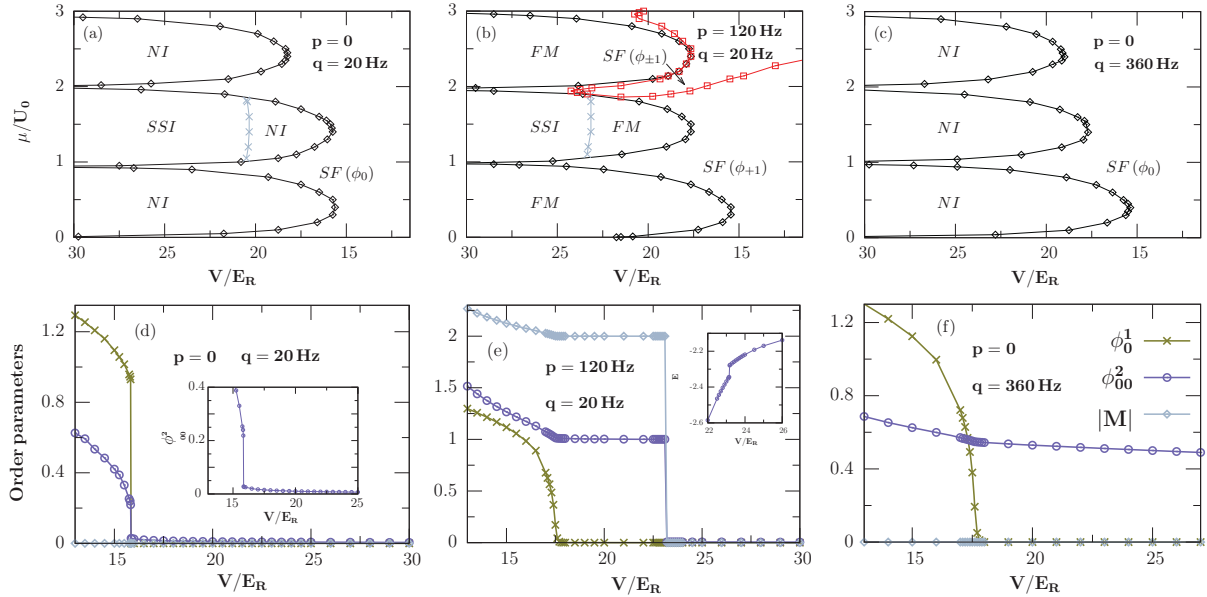}
\vspace{-4mm}
\caption{(Color online)  Zero-temperature phase diagrams (a),(b),(c) and phase transitions (d),(e),(f) of $^{23}{\rm Na}$ in a 3D optical lattice. Our calculations show a first- (second-) order SF-MI phase transition for weak [(d),(e)] (strong [(f)]) magnetic field with $\mu/U_0=1.5$. Inset: zoom of the panel (d) at the critical point of nematic order, and local energy around first-order phase transition within BDMFT for the panel (e), respectively.}\label{order}
\end{figure*}

\paragraph{Hopping dependent phase diagram.}

Away from the deep Mott-insulating regime, quantum fluctuations should be more and more important with the increase of tunneling amplitudes, and the stability of magnetic phases within the Mott lobes, such as the NI and SSI phase, is needed to be addressed against quantum fluctuations in the presence of the external magnetic field. We remark here that our results correctly recover unconventional spin ordering both in the atomic limit $U_0/t=\infty$ and in the weakly interacting regime. For even filling, our calculations prove that the SF-MI transition is first order for small external magnetic field (equivalent to small magnetism $M_z$ and small quadratic Zeeman interaction $q$), while it is second order for large magnetic field, as shown in Fig.~\ref{order}. We believe that the spin-dependent interaction $U_2$ will lead to the formation of spin-singlet pairs in the SSI phase which supports a first-order phase transition, while disappearance of the singlet pairs, due to the interplay of spin-dependent and Zeeman interactions, demonstrates a second-order transition.
Remarkably, we notice four distinctive types of SF phases in the parameters studied here, characterized by $\phi^{1}_{0}\neq0$, $\phi_{1}^{1}\neq0$ ($\phi^{1}_{-1}\neq0$), $\phi^{1}_{\pm1}\neq0$ or $\phi^{1}_{1,0}\neq0$, which emerge in the phase diagrams for different external linear Zeeman fields $p$ (magnetization $M_z$) and quadratic Zeeman fields $q$.
%Note here that the predicted various types of SF phases are relevant to finite-temperature Bose-Einstein condensation, and the corresponding details of multi-step condensation will be investigated in Fig.~\ref{multi_step}.
%BDMFT also captures the physics in the intermediate coupling regime (around the vicinity of the Mott transition), and predicts that insulating and strongly correlated superfluid phases have zero values of local magnetization $\bf {M}$ as well., even though it is relevant to directly observe these phases in realistic experiments

The order parameters as a function of interaction strengths are presented in Fig.~\ref{order}(d),(e),(f) for $U_2/U_0 = 0.037$, $\mu/U_0=1.5$, and different external magnetic fields. For zero magnetization $M_z$ and small quadratic Zeeman interaction, we find that a small region of the SSI phase ($n=2$) is occupied by the NI phase around the tip of the MI-SF transition as a result of the interplay of tunneling and spin-dependent interaction, as shown in the inset of Fig.~\ref{order}(d), where the corresponding MI-SF transition is first order. We remark here that the nematic order $\phi^2_{\alpha\beta}$ demonstrates a small but nonzero value in between SF and SSI phases, as shown in the inset of Fig.~\ref{order}(d), which indicates a possible nematic phase but may be an artifact of the BDMFT method, requiring further studies based on more accurate methods such as quantum Monte-Carlo simulations.  While for large quadratic Zeeman interaction, we observe that the spin-order in MI ($n=2$) is the NI type with $\phi^{1}_{\alpha}= 0$ and $\phi^2_{\alpha\beta}\neq0$, instead of SSI, and the corresponding transition is second order, as shown in Fig.~\ref{order}(f). Interestingly, for small but nonzero magnetization $M_z\neq 0$, we observe a FM phase with $\phi^{1}_{\alpha}= 0$ and ${\bf M}\neq 0$ in between SF and SSI, as shown in Fig.~\ref{order}(e). Interestingly, we observe a first-order SSI-FM and a second-order FM-SF transition, with lowering the optical depth $V$.

%Finally, in weakly interacting systems with $t\gg U_{0},U_{2}$, the expectation values of the spin components are zero along any direction~\cite{Tsuchiya,Lewenstein}, {\it i.e.} a polar ordering with $\bf {M} =0$, which is also predicted in our simulations. For relatively large hopping, we find a superfluid phase characterized by $\phi^1_\pm \neq 0$ ($\bf {M} =0$).Finally, in the weakly interacting regime with $t\gg U_{0},U_{2}$, a polar phase with zero magnetization $\mathrm{\mathbf{M}}=0$ is found in our simulations, which is consistent with findings in previous works~\cite{Tsuchiya,Lewenstein}.

\subsubsection*{Finite temperature}\label{section:finite}
\begin{figure}[t!]
\centering
\begin{tabular}{c}
\includegraphics*[width=0.6\linewidth]{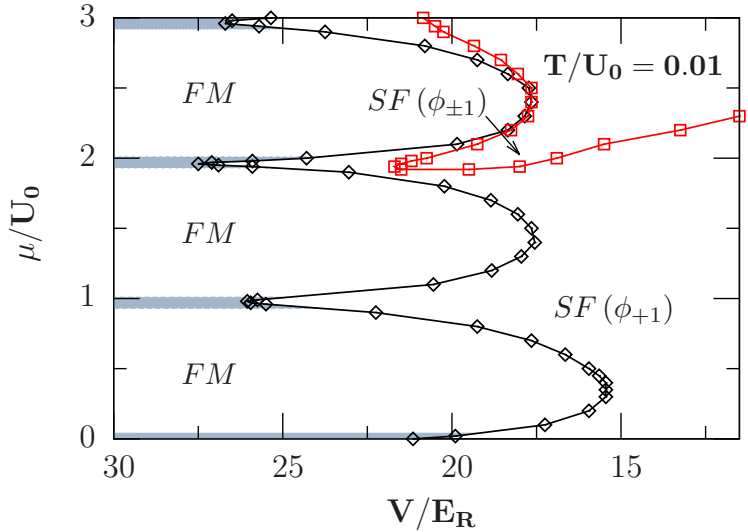}
\end{tabular}
\vspace{-3mm}
\caption{(Color online) Finite-temperature phase diagram for spin-1 ultracold bosons in a 3D cubic lattice with an antiferromagnetic interaction $U_2/U_0=0.037$ for a fixed temperature $T/U_0=0.01$, and a magnetic field $p=20\, Hz$ and $q=120\, Hz$. Note here that the normal state is shown by the shaded region in between FM.}\label{finite_T}
\end{figure}
\begin{figure}[t!]
\centering
\begin{tabular}{c}
\includegraphics*[width=0.9\linewidth]{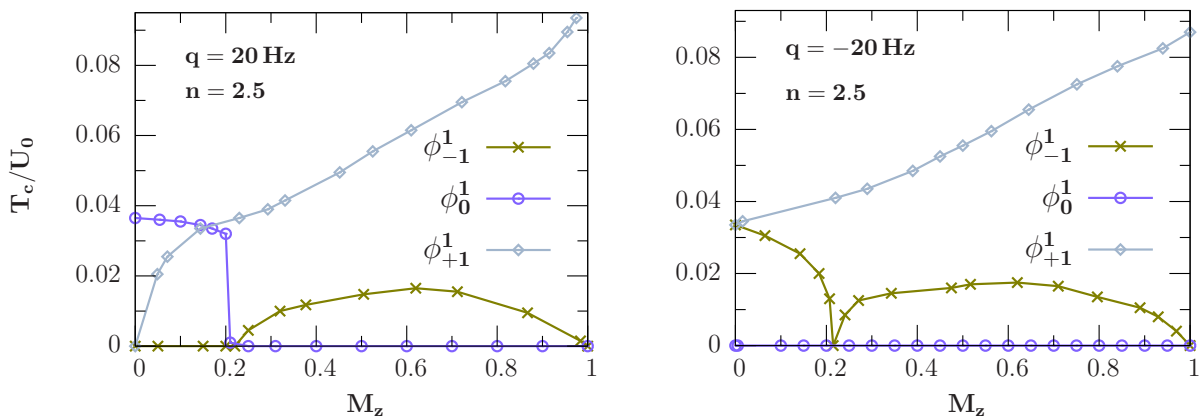}
\end{tabular}
\vspace{-5mm}
\caption{(Color online) Abnormal multi-step condensation of spin-1 ultracold bosons in a 3D cubic lattice ($V=21\,{\rm Hz}$) with an antiferromagnetic interaction $U_2/U_0=0.037$ and filling $n=2.5$ for quadratic Zeeman field $q=20\, Hz$ (left), and $-20\, Hz$ (right), respectively. Contrary to weakly interacting gases, we observe a non-smooth change of the critical condensing temperature of the $m_F=-1$ component.}\label{multi_step}
\end{figure}

Stability of the above spin-dependent phases against thermal fluctuations is the first thing to consider in order to obtain a direct observation of them in practical experiments. We investigate this issue for a typical case with $U_2/U_0 = 0.037$ ($^{23}\rm Na$) at a fixed finite temperature $T/U_0=0.01$,  and find that four different phases are involved as shown in Fig.~\ref{finite_T}, including SF ($\phi^1_{0}$ or $\phi^1_{\pm 1}$), FM, and normal state (NS) characterized both by $\phi^1_\alpha = 0$ and large density fluctuations $\Delta^2\equiv \langle n^2\rangle - \langle n\rangle ^2$. Due to thermal fluctuations, we find that the singlet pairs in $n=2$ are broken and the FM phase develops in the parameters studied here. Moreover, BDMFT predicts that the superfluid-Mott-insulating phase transition for even fillings is second order, and the transition between the SF phases ($\phi_{\pm}$ and $\phi_{+1}$) is second order as well. Notice that the temperature in our calculations can be obtained via present cooling schemes, for instance, the spin-gradient cooling~\cite{Weld2009,PhysRevA.92.041602}, and an external harmonic trap can be employed to maintain the coexistence of these phases.

\begin{figure}[t!]
\begin{tabular}{c}
\includegraphics*[width=0.9\linewidth]{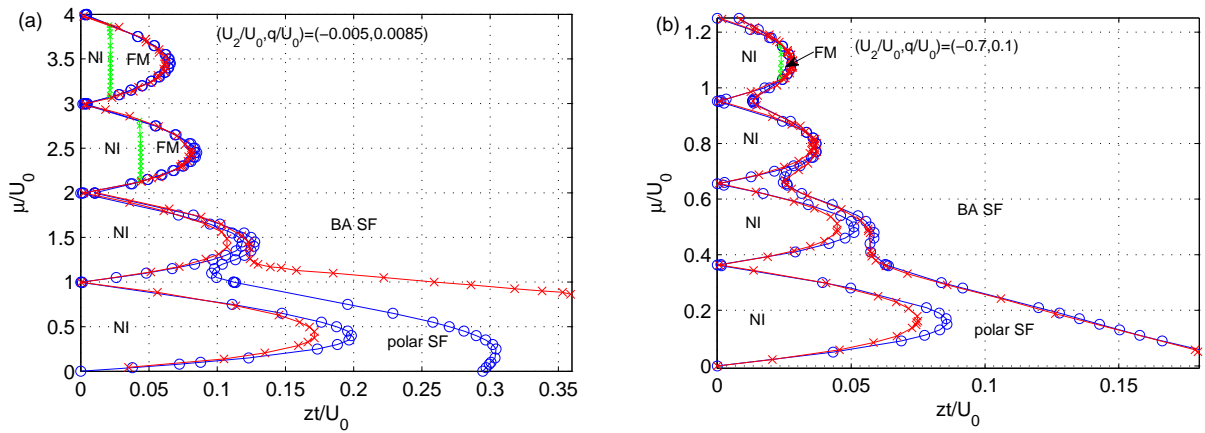}
\end{tabular}
\caption{\label{fig:1} (Color online) Zero-temperature phase diagrams of spin-1 ultracold bosonic gases in a 3D optical lattice for (a) $(U_{2}/U_{0},\,q/U_{0})=(-0.005,\,0.0085)$ ($^{87}$Rb) and (b) $(U_{2}/U_{0},\,q/U_{0})=(-0.7,\,0.1)$ ($^{7}$Li) respectively by BDMFT (blue circle and green cross), and via Gutzwiller mean-field theory in Ref.~\cite{PhysRevA.96.023628} (red cross). BDMFT predicts four phases in the system, including nematic insulator (NI), ferromagnetic (FM), polar and broken-axisymmetry (BA) superfluid (SF) phase.
}
\end{figure}

Another crucial question regarding experimental observations is the multi-step condensation of strongly correlated superfluid. For weakly interacting spinor gases, a two-step condensation with $m_F=1$ and $m_F = 0$ has been observed for large quadratic Zeeman energy, while there exists a three-step condensation for small quadratic Zeeman energy~\cite{zhang2003,Frapolli2017Stepwise}, due to the interplay of spin-dependent and Zeeman interactions. However, the condensation of strongly correlated superfluid is still unknown. Fig.~\ref{multi_step} shows the finite-temperature phase diagram of the spinor bosonic gases at fixed filling $n=2.5$ and quadratic Zeeman interaction $q=20\, {\rm Hz}$ (left) and $q=-20\, {\rm Hz}$ (right), respectively. An interesting feature is revealed in the neighbouring of the superfluid to Mott-insulating transition with $V=21\, {\rm Hz}$ ($t/U_0\approx0.01$). For example, for positive quadratic Zeeman field and small $M_z$, majority component $m_F = 0$ condenses first at a critical temperature, followed by the $m_F = 1$ component at a lower one, while, for large $M_z$, coexisting the $m_F = \pm 1$ components is energetically favored. This behavior is a result of the interplay of spin-dependent and quadratic Zeeman interactions (favor the $pair$ of the $m_F=\pm1$ components), and longitudinal magnetization (favor the $m_F=1$ component). Our prediction here is consistent with weakly interacting condensate, where a phase transition occurs between the broken-axisymmetry ($\phi^1_{0,1}\neq0$) and the antiferromagnetic phase ($\phi^1_{\pm 1} \neq 0$) for low temperature~\cite{zhang2003}, even though here we only observe a two-step condensation for small magnetization.
Interestingly, for negative quadratic Zeeman interaction, we observe an abnormal transition from the normal to the superfluid phase via cooling the system, $\it i.e.$  the critical condensing temperature of the $m_F=-1$ component first decreases into zero and then demonstrates an abnormal increase with magnetization $M_z$, which is a remarkable phenomenon never been predicted ever before, indicating unique features of strongly correlated superfluid.
%Actually, the spinor lattice bosons with ferromagnetic interactions also demonstrate abnormal features for condensing temperatures, and the corresponding details can be found in Fig.~\ref{fig:4} and related discussion..

In order to offer quantitative guidance for the direct observation of multi-step condensations in realistic experiments, we estimate the critical temperatures. For experiments with $^{23}{\rm Na}$ ($U_{2}/U_{0}\approx0.037$) in a 3D cubic lattice generated by laser beams of wavelength 1064 nm and intensity $V\approx21\, E_{R}$, we find that the system should be cooled down to $T_c\approx2\, {\rm nK}$ ($n=2.5$, $p=\pm20\, {\rm Hz}$). One can employ the excitation spectra or density correlations~\cite{Imambekov,Natu2015} to reflect these correlated insulating states by occupying, for example, Bragg scattering~\cite{Imambekov}, quantum gas microscopy~\cite{Gericke2008High,Bakr2009A,Sherson2010Single} or optical birefringence~\cite{Natu2015}, as long as the lifetimes of these states can meet the requirement of observation~\cite{Zhao2015}. Recently, a steady state of spinor bosons in optical lattices with a sufficiently long lifetime was reported~\cite{Jiang2016First}, as well as the existence of spin-nematic ordering in a spherical trap~\cite{PhysRevA.93.023614}.
\subsection*{Ferromagnetic interactions }\label{subsec:ferromagnetic interactions}
In this section, we mainly investigate experimentally accessible systems for $^{87}{\rm Rb}$ with a ferromagnetic interaction $U_2/U_0=-0.005$ and for $^{7}{\rm Li}$ with $U_2/U_0=-0.7$. Actually, the magnetic order is the trivial ferromagnetic order for zero magnetic field. At finite temperature, interestingly, the strongly correlated system may demonstrate a multistep condensation for different species, due to the interplay of quantum and thermal fluctuations.
\subsubsection*{Zero-temperature phase diagrams}
\begin{figure}[t!]
\begin{tabular}{c}
\includegraphics*[width=0.9\linewidth]{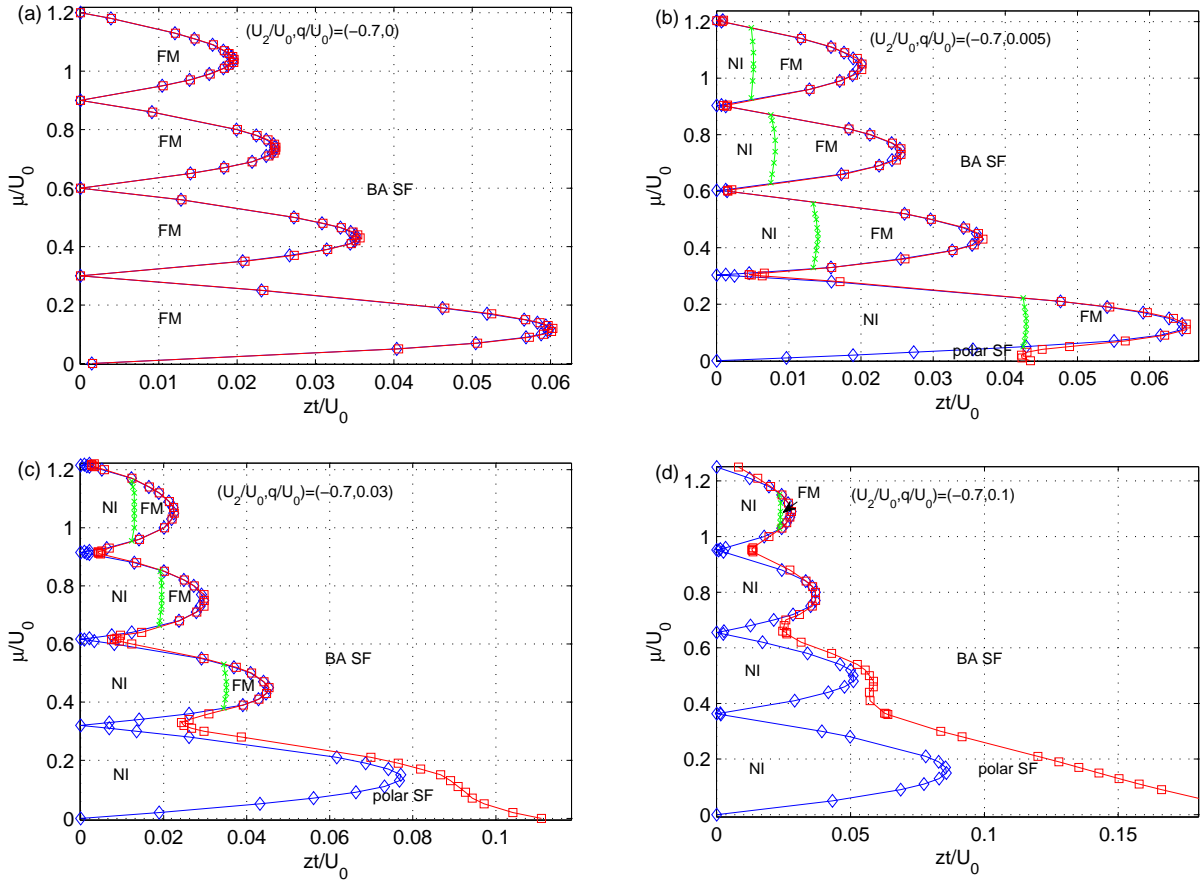}
\end{tabular}
\caption{\label{fig:2} (Color online) Influence of quadratic Zeeman interactions on zero-temperature phase diagrams of spinor bosonic gases in a 3D optical lattice. We choose the $^{7}$Li atom as examples with $(U_2/U_0=-0.7)$, and the other parameters are set to (a) $q/U_{0}=0$, (b) $q/U_{0}=0.005$, (c) $q/U_{0}=0.03$ and (d) $q/U_{0}=0.1$. Inside the Mott lobes, a competition between nematic insulating and ferromagnetic order is observed, as a result of the interplay of spin-dependent and quadratic Zeeman interactions.}
%FM phase domains the whole lobe when the quadratic Zeeman energy q is zero, while increasing q, NI phase begins to appear and squeeze the space of the FM phase. FM phase constantly decrease and finally almost disappear in panel (d).}
\end{figure}
In this subsection, we discuss zero-temperature phase diagrams of spinor ultracold gases in a 3D optical lattice. For comparison, we firstly study the cases in Ref.~\cite{PhysRevA.96.023628}, where a Gutzwiller mean-field calculation is performed for different interactions. Fig.~\ref{fig:1} shows the obtained phase diagrams for $(U_{2}/U_{0},q/U_{0})=(-0.7,0.1)$ for $^7$Li  and $(-0.005,0.0085)$ for $^{87}$Rb, based on BDMFT. Here, we notice that there exists four distinct phases obtained from BDMFT, namely polar superfluid (polar SF), broken-axisymmetry superfluid (BA SF), nematic insulator (NI) and ferromagnetic (FM). In the superfluid phase near to the Mott lobes, only the $\sigma\!=\!0$ component shows superfluidity, which we identify as the polar superfluid phase.
The other superfluid phase has three superfluid components, and possesses a non-zero transverse magnetization $M_{tr} := \sqrt{\langle S_{(x)}\rangle^2+\langle S_{(y)}\rangle^2}$; hence we identify it as the broken-axisymmetry superfluid phase.
The nematic insulator phase is characterized by $\phi^{1}_{\alpha}=0,\phi^{2}_{\alpha\beta}>0$ and $\bm{M}=0$; and the ferromagnetic phase for $\phi^{1}_{\alpha}=0$ and $\bm{M}\neq0$.
%BDMFT enables us to distinguish the phases and the phase transition boundary inside the Mott lobe as shown in Fig. \ref{fig:1}.
We remark here that our method clearly resolves the long-range spin order. As shown in Fig.~\ref{fig:1}(a), there are two different Mott-insulating phases inside the $n=3$ and $4$ Mott lobes, $\it i.e.$ the NI phase for small $zt/U_{0}$ and the FM phase for large $zt/U_{0}$, while, for $n=1$ and $2$, the Mott-insulating phase favors the nematic order, as a result of the interplay of spin-dependent and quadratic Zeeman interactions.
\begin{figure}[t!]
\centering
\includegraphics[width=0.75\linewidth]{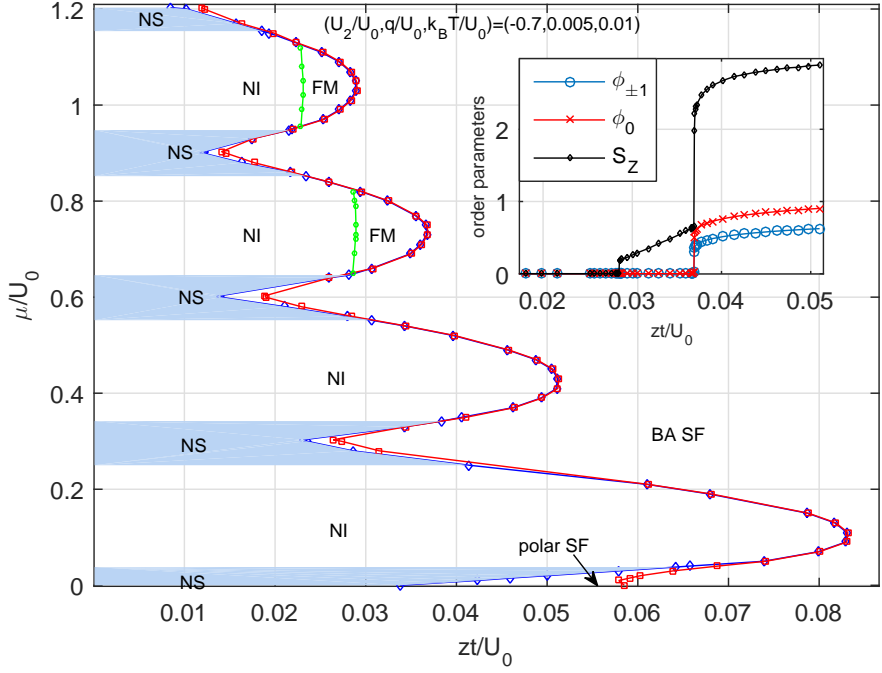}% Here is how to import EPS art
\caption{\label{fig:3} (Color online) Influence of thermal fluctuations on phase diagram of spinor bosonic gases in a 3D optical lattice for $(U_{2}/U_{0},q/U_{0},T/U_{0})=(-0.7,0.005,0.01)$ ($^{7}$Li). The diagram manifests five different phases: polar superfluid (polar SF), broken-axisymmetry superfluid (BA SF), nematic insulator (NI), ferromagnetic phase(FM), and normal state (NS). We observe a second-order NI-SF ($n=1$ and $2$), and a first-order NI-FM-SF ($n=3$ and $4$) phase transition, respectively, as shown by the inset with $\mu/U_0=0.72$. }
%We obtain that with the existence of temperature, the boundary of the SF phase transition has a rightward translation. Besides, the minimum values of the lobe are nonzero, NI phase overweights FM phase and begins to domain the whole area inside the lobe.
\end{figure}

Next, we compare our results with those obtained from Gutzwiller mean-field theory in Ref.~\cite{PhysRevA.96.023628}. For example, both methods predict a MI-polar SF and a polar SF-BA SF phase transition. We also find that phase boundaries are well matched with previous results, except that the Mott tip here is slightly bigger than that from the Gutzwiller variational ansatz, since that BDMFT takes quantum fluctuations into account. However, two major differences are also found between the two theories. First, magnetic spin orders of the Mott-insulating phases are resolved via BDMFT, which includes nematic order and ferromagnetic order in the parameters studied here. Second, Gutwziller mean-field theory underestimates the stabilized region of the polar SF phase for $^{87}$Rb atoms, $\it i.e.$ there is an obvious mismatch of the polar SF-BA SF transition boundary. This discrepancy is expected to be due to the lack of quantum fluctuations involved in the Gutzwiller variational ansatz.

To figure out the relationship and competition between the long-range spin orders, we focus on a $^{7}$Li gas with spin-dependent interaction $U_2/U_0=-0.7$, and investigate the influence of quadratic Zeeman interactions. It is expected that the NI phase appears in the Mott-insulating lobe and occupies the region of the FM phase with increasing $q$, as a result of the interplay of spin-dependent and quadratic Zeeman interactions, which supports the $m_F=1$ and the $m_F=\pm 1$ components, respectively. The corresponding phase diagrams for different parameters are obtained and summarized in Fig. \ref{fig:2}. As examples, we choose $q/U_{0}=0,0.005,0.03$ and $0.1$, respectively. When $q/U_{0}=0$, the Mott lobes are occupied by the FM phase, which is consistent with previous conclusions~\cite{Katsura2013}, while there is only broken-axisymmetry superfluid phase outside the Mott lobe. With the increasing of quadratic Zeeman energy $q/U_{0}$, the region of the MI phase gradually widen with the appearance of the NI order from the lower hopping regime, as shown in Fig.~\ref{fig:2}(b), where the polar SF phase emerges at $q/U_{0}=0.005$. Increasing $q$ further, the polar SF phase expands and squeezes the region of the BA SF phase, as shown in Fig.~\ref{fig:2}(b),(c),(d). Inside the Mott lobes, a competition between the NI and the FM order is observed, $\it i.e.$ the region of the NI phase expands larger and the region of the FM phase shrinks with the increasing of $q/U_{0}$.
Finally, the FM phase disappears at $(U_{2}/U_{0},q/U_{0})=(-0.7,0.1)$, as shown in Fig.~\ref{fig:2}(d).

%The phase diagram exhibits saturated ferromagnetism when the $q/U_{0}$ is zero, the results are consistent with theoretical study\cite{Katsura2013}.

\subsubsection*{Finite temperature phase diagrams}
Up to now, we have illustrated the zero-temperature case and studied quantum-fluctuation-induced phase transitions. In this subsection, we investigate the influence of thermal fluctuations on phase diagrams of spinor ultracold gases in an optical lattice with ferromagnetic interactions. We take $(U_{2}/U_{0},q/U_{0})=(-0.7,0.005)$ as an example. For a typical parameter with $T/U_{0}=0.01$, the finite-temperature phase diagram is shown in Fig. \ref{fig:3}. We observe there are five phases in the system, including NI, FM, BA SF, polar SF, and normal state (NS) characterized both by $\phi^1_\alpha = 0$ and large density fluctuations $\Delta^2\equiv \langle n^2\rangle - \langle n\rangle ^2$.
Compared with the zero-temperature case, the nonzero-temperature one possesses some features worth mentioning.
While the SF-MI phase boundary expands, the Mott-insulating lobes show major changes. For example, the $n=1$ Mott lobe exhibits the saturated NI phase and the FM phase disappears completely, since thermal fluctuations smooth the population difference between the three species and the FM phase is energetically unfavored. Similarly, for the Mott lobes with $n=2,3,4$, the NI phase expands and squeeze the rest region of the FM phase. We notice some odd behaviors relating to NI-FM phase boundaries for $n=3,4$, while the reasons behind this abnormal behavior remain unknown.

\begin{figure}[ht]
\begin{tabular}{c}
\includegraphics*[width=0.9\linewidth]{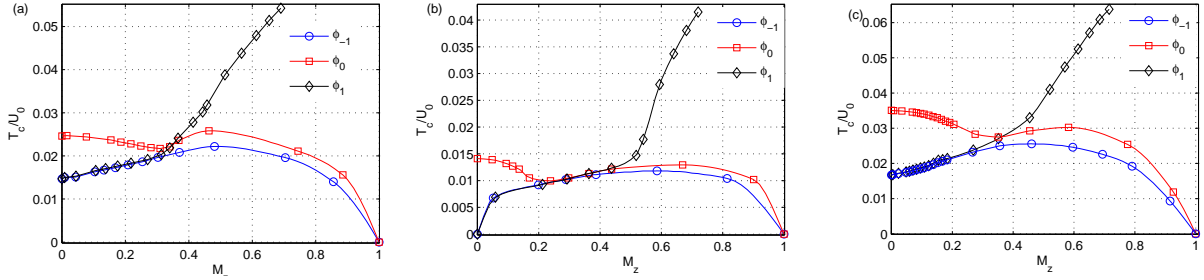}
\end{tabular}
\caption{\label{fig:4} (Color online) Abnormal multistep condensation for spin-1 ultracold bosons in a 3D cubic lattice with a ferromagnetic interaction $U_2/U_0=-0.7$ ($^{7}$Li) for fixed filling $n=1$ (a),(b) and $n=2$ (c). The other parameters are set to $zt/U_0=0.096$ (a) and $0.084$ (b), and $zt/U_{0}=0.063$ and $q/U_{0}=0.1$ (c). With increasing $M_z$, the system demonstrates interesting features, where a local minimum appears for the critical condensing temperature of the $m_F=0$ component.}
\end{figure}

One crucial question regarding experimental observations is the multi-step condensation of strongly correlated superfluid. While the connection between magnetic order and Bose-Einstein condensation of weakly interacting spin-1 bosons with ferromagnetic interactions is debated~\cite{RevModPhys.85.1191}, the investigation on multistep condensations of strongly correlated lattice bosons is still missing. Here, we focus on the equilibrium states of the strongly interacting superfluids near the Mott lobes under the constraint of constant longitudinal magnetization (in the absence of dipolar relaxation, longitudinal magnetization is constant, instead of a constant magnetic field).
We examine how an applied external magnetic field affects the critical condensing temperature $T_{c}$. In our calculations, we fix the value of the hopping matrix element $zt/U_{0}$ and filling number of per site $n$, and obtain the relationships between $T_{c}$ and the local magnetization $M_{z}$, as shown in Fig.~\ref{fig:4}. In Fig.~\ref{fig:4}(a),(b), we choose $(U_{2}/U_{0},q/U_{0})=(-0.7,0.03)$, $n=1$, and $zt/U_{0}=0.096$ and $0.084$, respectively. The selected hopping amplitudes have unique properties, where $zt/U_0=0.096$ locates in the BA SF phase at zero magnetization with $\phi^{1}_{0,\pm1}\neq0$, and $zt/U_0=0.084$ sits in the polar SF phase with $\phi_0\neq0$. In Fig~\ref{fig:2}(c), we focus on another case with large filling $n=2$.

We find that the system demonstrates interesting features as a function of longitudinal magnetization. With increasing $M_z$, the critical condensing  temperature of the $m_F=+1$ component increase monotonously, which is consistent with weakly interacting case~\cite{RevModPhys.85.1191}, since this component is energetically favored. However, the critical temperature of the $m_F=-1$ component increases at first and then decreases to zero, while the $m_F=0$ component decreases initially, then increases, and finally decreases to zero, which is inconsistent with weakly interacting case~\cite{RevModPhys.85.1191}. The abnormal changes of condensing temperatures are the results of the interplay of spin-dependent interactions and quadratic Zeeman effect, and longitudinal magnetization, where the population of the $m_F=+1$ component increases with longitudinal magnetization, and quadratic Zeeman interactions favors the "pair" of the $m_F=\pm1$ component. Surprisingly, we observe a local minimum of the critical temperature of the $m_F=0$ component, which is a phenomenon has never been observed ever before. To obtain a better understanding of this behavior, we plot the multistep condensation of the spinor gases for another parameters, as shown in Fig.~\ref{fig:4}(c), with $(U_{2}/U_{0}, q/U_{0})=(-0.7,0.1)$, $n=2$, and $zt/U_{0}=0.063$. We observe a minimum of the critical temperature of the $m_F=0$ component. We remark that the spinor lattice bosons with antiferromagnetic interactions also demonstrate abnormal features for critical condensing temperatures.

\section*{Discussion}
In conclusion, we employ spinor bosonic dynamical mean-field theory to carry out extensive calculations of ultracold spinor Bose gases loaded into a cubic optical lattice. Complete phase diagrams of the system with both antiferromagnetic and ferromagnetic interactions are obtained. Various phases, including nematic, ferromagnetic and spin-singlet insulator, polar superfluid, and broken-axisymmetry superfluid, are found. In particular, the competition between above phases are investigated by varying related physical influences such as spin-dependent interactions, quadratic Zeeman energy, and thermal fluctuations. Multistep condensations of the strongly correlated superfluids are explored as a function of longitudinal magnetization and temperature. Interestingly, the critical temperature of the $m_F=-1$ component increases firstly and then decreases to zero with increasing $M$ for antiferromagnetic interactions, while the critical temperature of the $m_F=0$ component demonstrates a local minimum for ferromagnetic cases, which is inconsistent with weakly interacting spinor gases.

\bibliography{apstemplate}

\section*{ACKNOWLEDGMENTS}
We acknowledge useful discussions with Y.-M. Liu, J.-M. Yuan, Z.-X. Zhao.
This work was supported by the National Natural Science Foundation of China under Grants No. 11304386, 11774428, and No. 11104350.
\section*{Author Contributions Statement}
J.W. and Y.L. designed the study. X.Z. and J.L. performed calculations. J.H. analyzed the data. X.Z. and Y.L. wrote and reformulated the manuscript. All authors discussed the results and contributed to the manuscript.

\section*{Additional information}
\textbf{Competing interests:} The authors declare no competing interests.

%\noindent Correspondence and requests for material should be sent to li\_yq@nudt.edu.cn.

\end{document}